\documentclass[a4paper]{article}

\usepackage{INTERSPEECH2022}
\usepackage{algorithm}
\usepackage{algpseudocode}
\usepackage{multirow}
\usepackage{array}
\usepackage{url}
\usepackage{cite}
\usepackage{bm}

\title{EDITnet: A Lightweight Network for Unsupervised Domain Adaptation in Speaker Verification}
\name{Jingyu Li, Wei Liu, Tan Lee}
\address{Department of Electronic Engineering, The Chinese University of Hong Kong, Hong Kong}
\email{\{lijingyu0125, louislau\_1129\}@link.cuhk.edu.hk, tanlee@ee.cuhk.edu.hk}

\begin{document}

\maketitle
\begin{abstract}
Performance degradation caused by language mismatch is a common problem when applying a speaker verification system on speech data in different languages. This paper proposes a domain transfer network, named EDITnet, to alleviate the language-mismatch problem on speaker embeddings without requiring speaker labels. The network leverages a conditional variational auto-encoder to transfer embeddings from the target domain into the source domain. A self-supervised learning strategy is imposed on the transferred embeddings so as to increase the cosine distance between embeddings from different speakers. In the training process of the EDITnet, the embedding extraction model is fixed without fine-tuning, which renders the training efficient and low-cost. Experiments on Voxceleb and CN-Celeb show that the embeddings transferred by EDITnet outperform the un-transferred ones by around $30\%$ with the ECAPA-TDNN512. Performance improvement can also be achieved with other embedding extraction models, e.g., TDNN, SE-ResNet34.

\end{abstract}
\noindent\textbf{Index Terms}: speaker verification, domain adaptation, conditional variational auto-encoder, self-supervised learning

\section{Introduction}
\label{sec:Introduction}
Speaker verification (SV) is to determine whether an input utterance is from a claimed speaker. Recently, deep neural network (DNN) based models have shown great success on SV \cite{liu2015deep,variani2014deep,snyder2018x,snyder2017deep,xie2019utterance,chung20b_interspeech,desplanques20_interspeech}. Nevertheless, DNN models are known to have robustness problem against domain mismatch (DM) \cite{sadjadi20172016}. 
%With large mismatch of the model's training data and the test data, the model performance is commonly declined severely(). 
As human speech contains a diverse range of information, domain mismatch in SV could be attributed to multiple factors, namely, background noise, channel distortion, and language difference. 
%Language mismatch is regarded as one of the most common DM which has large impact on SV and a novel domain adaption(DA) is proposed in this work to alleviate this problem. 
% In practical applications, the speakers may use different languages. The SV system trained by the data from one language will encounter the mismatch problem when being utilized in another language, and we focus on the language mismatch in this paper. 
% An SV system trained by speech data from one language would encounter serious mismatching problem when being applied to speech input in another language. 
Language mismatch in the training and test data raises a big challenge in building a good SV system\cite{sadjadi20222021}.
This is common in practical applications because languages used by speakers are unconstrained. In this paper, we investigate methods of domain adaptation (DA) to suppress the effect of language mismatch in SV. Specifically, if an SV model is trained on language $\mathcal{A}$ and adapted to language $\mathcal{B}$,  $\mathcal{A}$ and $\mathcal{B}$ are referred to as the source domain and the target domain respectively. 

Numerous DA methods were proposed in the field of SV. Regarding whether or not speaker labels are required for the target-domain speech data, the methods are categorized as supervised DA and unsupervised DA. Supervised DA works by fine-tuning the source-domain model with target-domain data with speaker labels\cite{fan2020cn}. Fine-tuning embedding extraction model is time-consuming and may have the issue of over-fitting if data amount is limited. In many cases, speaker labels may not be available, e.g., for low-resourced languages. Here we focus on unsupervised DA since it is more relevant to real-world application scenarios. 

% Related work: the purpose of DA, method...
% The goal of domain adaptation is to pull the source and target domain distribution together as close as possible. 
Two kinds of approaches for DA are commonly adopted to reduce the difference in the domain distributions between target and source. One is to impose the source domain information on the target data, which can be done by explicitly matching the target domain's feature distribution to the source domain's one, in terms of the first-order or second-order statistics. For example, Correlation alignment (CORAL) 
% is a fast unsupervised adaptation approach, that 
aims to minimize the distance of co-variance between the two domains \cite{alam2018speaker}. Maximum mean discrepancy (MMD) 
% is widely applied to 
reduces domain mismatch by minimizing the mean-squared difference of the statistics of features from different domains\cite{sun2016deep, lin2020framework, wang2020cross}. 

Another strand of approaches is to reduce the domains' specific information implicitly by Domain Adversarial Training (DAT)\cite{ganin2015unsupervised}. In DAT, a domain critic is trained jointly in an adversarial manner, in addition to the original speaker classification training in the source domain. 
%The purpose of adversarial training is to make the extracted embeddings to be speaker discriminative and domain invariant . 
The min-max optimization of DAT enforces the two domain distributions to be close to each other so that the extracted embeddings become speaker-discriminative and domain-invariant \cite{wang2018unsupervised, xia2019cross, rohdin2019speaker, chen2020adversarial}. In \cite{tu2019variational, tu2020variational}, it was shown that producing Gaussian distributed speaker representations helped alleviate the mismatch and was amenable to the PLDA backend. Inspired by the recent progress of self-supervised representation learning, a self-supervised based domain adaptation method (SSDA) was proposed in \cite{chen2021self} to fully leverage available information from the target-domain data.

% motivation
% Although these adaptation methods are effective, most of them are required to jointly fine-tune with the backbone of SV, that is the heavy speaker embedding extractor. 
In the present study, a lightweight network, named Embedding Domain Inter-Transfer Network (EDITnet), is proposed for unsupervised DA. 
The network is applied to speaker embeddings extracted by an embedding extraction model, which is pre-trained on source-domain data and fixed in the subsequent processes. 
% The network is applied at the speaker embedding level, in which the embedding extraction model is pre-trained on the source domain and fixed in the training process of EDITnet. 
EDITnet uses a Conditional Variational Auto-encoder (CVAE) to transfer speaker embeddings from the target domain into source domain. Moreover, a self-supervised loss is imposed on the transferred output to increase the distance between speakers' embeddings for better speaker discrimination. Free from fine-tuning, EDITnet can achieve fast and light implementation in both training and evaluation stages. It is also suitable for SV without target-domain labels, and can be widely used for different languages. In our experiments, the proposed model is evaluated with different embedding extraction models and shows significant performance.

The rest of the paper is organized as follows. Section 2 briefly describes the CVAE and then introduces the architecture of the proposed model. The experimental setup is illustrated in Section 3. Section 4 gives the detailed experimental results and the paper is concluded in Section 5.

\section{Methodology}
\label{sec:Methodology}
%The DM of input audio data brings mismatch into the speaker embedding generated from a DNN model, which decreases the accuracy of SV dramatically. 
%EDITnet is proposed in this work to transfer the target embeddings into source domain, which alleviates the mismatch of embeddings from different languages and improve the verification results. 
% The proposed EDITnet utilizes a CVAE structure to transfer the target embeddings into the source domain, which alleviates the mismatch of embeddings from different languages and improves the verification results. 
First, the basic ideas of CVAE are described, and the details of EDITnet are given in the following.

\subsection{Conditional VAE}
\label{ssec:CVAE}
% general CVAE
The variational auto-encoder (VAE) is a generative model, that poses the variational regularization on the latent space \cite{kingma2013auto}. VAE comprises two modules, namely encoder and decoder. 
% The encoder module converts the input $\mathbf{x}$ into a latent variable $\mathbf{z}$. The decoder maps $\mathbf{z}$ back to generate output $\hat{\mathbf{x}}$, which is supposed to be a reconstructed $\mathbf{x}$. That $\hat{\mathbf{x}}$ is expected to be similar to $\mathbf{x}$. 
The encoder module converts the input $\mathbf{x}$ into a latent variable $\mathbf{z}$ and the decoder maps $\mathbf{z}$ back to generate output $\hat{\mathbf{x}}$. $\hat{\mathbf{x}}$ is the reconstruction of $\mathbf{x}$ and is expected to be similar to $\mathbf{x}$. 
The prior distribution of $\mathbf{z}$, i.e., $p(\mathbf{z})$, is usually modeled by a multivariate normal distribution. In practice, the VAE model is trained to maximize the Evidence Lower Bound (ELBO) of the data log-likelihood $log\ p(\mathbf{x})$. The conditional VAE (CVAE) is a variant of VAE\cite{sohn2015learning}. The distribution of latent variable $\mathbf{z}$ in CVAE is conditioned on the label $\mathbf{c}$ as $p_\epsilon(\mathbf{z}|\mathbf{c})$, while the vanilla VAE maps all input data into the same latent space $p(\mathbf{z})$.
%, called conditional prior distribution. 
In this work, the condition label $\mathbf{c}$ specifies the domain category of the input data, i.e., target or source. The training objective of CVAE is to maximize the ELBO of conditional log-likelihood $log\ p(\mathbf{x}|\mathbf{c})$, which is written as follows:  
\begin{equation}
    \begin{aligned}
        \mathcal{L}_{CVAE}(\mathbf{x}, \mathbf{c}; \epsilon, \phi, \theta) =  
        & ~ \mathbb{E}_{q_{\phi}(\mathbf{z}|\mathbf{x},\mathbf{c})}[log\ p_{\theta}(\mathbf{x}|\mathbf{z},\mathbf{c})] - \\
        & KL(q_{\phi}(\mathbf{z}|\mathbf{x}, \mathbf{c})||p_\epsilon(\mathbf{z}|\mathbf{c})), 
    \end{aligned}
  \label{eq:CVAE}
\end{equation}
where $\epsilon$, $\phi$, and $\theta$ are the parameters of the conditional prior distribution, encoder, and decoder, respectively. The first term on the right-hand side represents how well the input $\mathbf{x}$ can be reconstructed and the second term measures how closely the latent code $\mathbf{z}$ follows the prior distribution. The two terms are denoted as $loss_{rec}$ and $loss_{KL}$ in the following sections.

By incorporating the condition label $\mathbf{c}$, CVAE is able to map data from various domains to different latent distributions and show better modeling ability than VAE\cite{sohn2015learning}.

\begin{figure}[t]
  \centering
  \includegraphics[width=\linewidth]{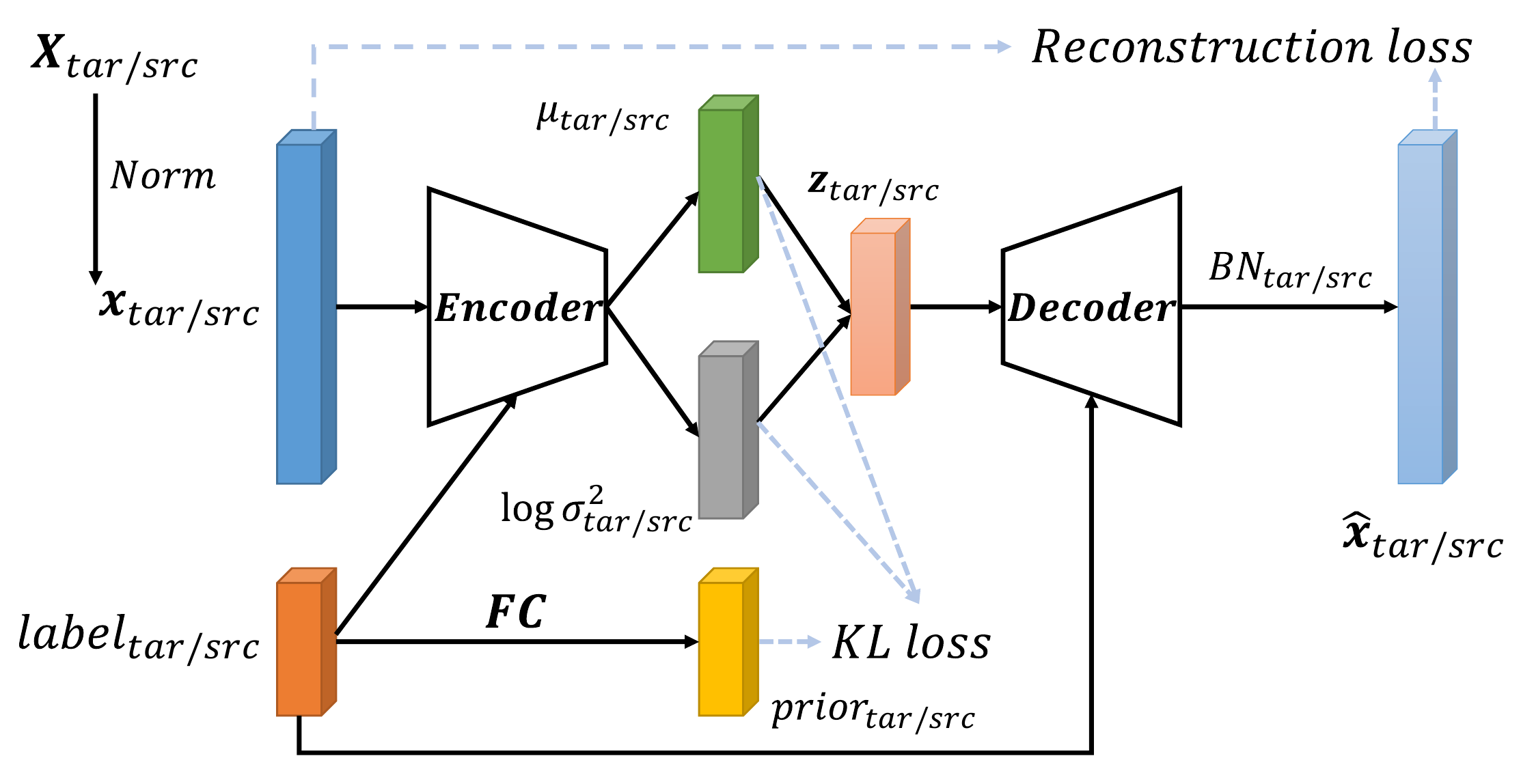}
  \caption{The diagram of the data flow. The target domain and source domain are represented as $tar$ and $src$ for short respectively.}
  \label{fig:CVAE}
\end{figure}

\subsection{Architecture of the EDITnet}
\label{ssec:archi}
The proposed EDITnet leverages a CVAE structure to perform domain transfer on speaker embeddings. The data flow pipeline of the model is shown in Fig.~\ref{fig:CVAE}, where $\mathbf{X}_{tar/src}$ denotes the extracted embeddings from the target or source domain. The embeddings are normalized before being fed into the encoder. $\mu$ and $log\  \sigma^2$ are computed from the encoder, representing the mean and logarithmic variance of the latent variable distribution. The latent variable $\mathbf{z}$ is sampled from $\mathcal{N}(\mu, \sigma^2)$. The embeddings from the target domain use a one-hot variable $[1, 0]$ as the conditional label $\mathbf{c}$, and $[0, 1]$ is utilized by the source domain. 
The prior distribution ($p_\epsilon(\mathbf{z}|\mathbf{c})$ in Eq.~\ref{eq:CVAE}) of two domains' latent variables are modeled by two normal distributions as $\mathcal{N}(prior_{tar}, I)$ and $\mathcal{N}(prior_{src}, I)$ respectively, where $I$ denotes the identity matrix. $prior_{tar/src}$ is given by a fully-connected layer (FC) in EDITnet (see Fig.~\ref{fig:CVAE}). 
% The latent variables of the two domains are modeled by two normal distributions as $\mathcal{N}(prior_{tar}, I)$ and $\mathcal{N}(prior_{src}, I)$. $prior_{tar/src}$ is given by the prior distribution ($p_\epsilon(\mathbf{z}|\mathbf{c})$ in Eq.~\ref{eq:CVAE}), which is a fully-connected layer (FC) in EDITnet (see Fig.~\ref{fig:CVAE}). 
At the output of the decoder, there are two batch normalization (BN) layers for the target domain and source domain respectively. The dimensions of the speaker embedding $\mathbf{x}$ and the latent variable $\mathbf{z}$ are $256$ and $128$. Details of the EDITnet model architecture are given in Table~\ref{tab:cvae_structure}.

\begin{table}
        \caption{The details of the CVAE structure used in this model. Concat. is short for concatenation.}
        \label{tab:cvae_structure}
        \small{
        \centering
        % \scalebox{1}{
        \renewcommand{\arraystretch}{1.2}
        \begin{tabular}{l|c|c}
        \toprule 
        \textbf{Block} & \textbf{Structure} & \textbf{in $\times$ out size} \\
        \hline
        \hline
         & Concat. of $\mathbf{x}$ and $label$  & $[256, 2] \times 258$ \\
        \cline{2-2}  \cline{3-3}  
         & [FC,ReLU,BN] & $258 \times 256$ \\
         \cline{2-2}  \cline{3-3}  
        Encoder  & FC & $256 \times 128$\\ 
         \cline{2-2}  \cline{3-3}  
        & tanh & $128 \times 128$ \\
        \cline{2-2}  \cline{3-3}  
         & $\mu$: FC, $log\ \sigma$: FC  & $128 \times 128$\\
         \hline 
         & Concat. of $\mathbf{z}$ and $label$  & $[128, 2] \times 130$ \\
        \cline{2-2}  \cline{3-3}  
         & [FC,ReLU,BN] & $130 \times 256$ \\
        \cline{2-2}  \cline{3-3}  
        Decoder & [FC,ReLU,BN]& $256 \times 512$  \\
        \cline{2-2}  \cline{3-3} 
        & FC & $512 \times 256$  \\
        \cline{2-2}  \cline{3-3}  
         & BN$_{tar/src}$  & $256 \times 256$  \\
        \hline 
        Prior & FC & $2 \times 128$ \\
        \bottomrule 
        \end{tabular}
        \renewcommand{\arraystretch}{1}
        % }
    }
\end{table}

In EDITnet, data from different domains are mapped onto different distributions in the latent variable space according to their conditional labels. $prior$ represents the center of the latent variable $\mathbf{z}$'s distribution, and we consider it contains the input domain information. Thus the variable transfer in the latent space is important in generating transferred embedding output. The embedding transfer process is given in Algorithm~\ref{alg:transfer}. Different from the reconstruction $\mathbf{\hat{x}}_{tar}$, the latent variable $\mathbf{z}_{tar}$ is shifted according to the discrepancy of the two $prior$s and decoded using the conditional label from the source domain.

\begin{algorithm}
\caption{Transfer embedding $\mathbf{x}_{tar}$ from target domain to source domain}
\label{alg:transfer}
\begin{algorithmic}[1]
\State $\mu_{tar}, log\ \sigma^2_{tar}=Encoder(\mathbf{x}_{tar}, label_{tar})$ 
\If{training}
    \State $\mathbf{z}_{tar} \sim \mathcal{N}(\mu_{tar}, \sigma_{tar}^2)$ \Comment{sampling}
\Else
    \State $\mathbf{z}_{tar}=\mu_{tar}$  \Comment{for evaluation}
\EndIf
\State $\mathbf{\widetilde{z}}_{tar}=\mathbf{z}_{tar}-prior_{tar}+prior_{src}$ \Comment{transfer the latent variable into source domain}
\State $\mathbf{\widetilde{x}}_{tar}=BN_{src}(Decoder(\mathbf{\widetilde{z}}_{tar}, label_{src}))$
\State\Return $\mathbf{\widetilde{x}}_{tar}$ as the transferred embedding of $\mathbf{x}_{tar}$
\end{algorithmic}
\end{algorithm}

The training loss of the model comprises three major parts. The first two losses come from the Eq.~\ref{eq:CVAE}. $loss_{rec}$ is given by the $L2$ distance between $\mathbf{\hat{x}_{tar/src}}$ and $\mathbf{x_{tar/src}}$. $loss_{KL}$ consists of the KL divergences from the two domains, i.e., $KL\{\mathcal{N}(\mu_{tar/src}, \sigma_{tar/src}^2),\  \mathcal{N}(prior_{tar/src}, I)\}$. They are calculated as follows:
\begin{equation}
  loss_{rec} = \frac{1}{N}\sum_{n=1}^N |\mathbf{x}_n -\mathbf{\hat{x}}_n|^2 
\label{eq:CVAE_loss_rec}
\end{equation}
\begin{equation}
\begin{aligned}
  loss_{KL} = -\frac{1}{N}\sum_{n=1}^N&\frac{1}{2}\sum_{j=1}^J (1 + log(\sigma^2_{n,j})- \\
              & (\mu_{n,j}-prior)^2 - \sigma_{n,j}^2) 
 \end{aligned}
\label{eq:CVAE_loss_kl}
\end{equation}
where $N$ is the number of samples and $J$ denotes the dimension of the latent variable $\mathbf{z}$. 

Self-supervised learning (SSL) has been widely applied to extract representative information with unlabeled data. The SSL cosine loss is applied in the training of EDITnet in order to push the transferred embeddings further away from each other in the cosine space. Denoting the cosine similarity between two embeddings as $cos\langle x,y \rangle$, the loss function is given as:
\begin{equation}
  loss_{cos(x,y)} = ReLU(-log(1-cos\langle x, y \rangle))
\label{eq:CVAE_loss_cos}
\end{equation}
A $ReLU$ activation is applied on the entropy function output, thus the loss is optimized only when $angle\langle x,y \rangle < 90^\circ$. The inputs are randomly sampled from training data and we assume that most of the samples in a mini-batch come from different speakers. The $loss_{cos}$ is applied between all $\mathbf{\widetilde{x}}_{tar}$ embeddings in a mini-batch except the embedding to itself. Besides, as the transferred embeddings share the same domain with $\mathbf{x}_{src}$, the cosine repulsion is also applied between $\mathbf{x}_{src}$ and $\mathbf{\widetilde{x}}_{tar}$.  

The total loss is given as the summation of $loss_{rec}$, $loss_{KL}$ and $loss_{cos}$. No supervised training, e.g., speaker classification, is required in the EDITnet.

% and $\langle \widetilde{X_{tar}}, X_{src} \rangle $.A cosine distance loss is applied on the transfer embeddings to push the embeddings far away from each other on the cosine space.

% $\widetilde{X_{tar}}$  The cosine loss in this stage is given as:
% \begin{equation}
%   loss_{cos} = \frac{1}{M\times N}\sum_{m,n}^{M,N} ReLU(-log(1-cos\langle X_{src}^{(m)}, \widetilde{X_{tar}^{(n)}}\rangle))
% \label{eq:CVAE_loss_cos}
% \end{equation}

% In the reconstruction of CVAE, $BN_{tar}$ and $BN_{src}$ are applied for $\hat{X_{tar}}$ and $\hat{X_{src}}$ respectively. 

% Thus the KL divergence of the latent variables are given by $KL\{\mathcal{N}(prior_{src}, I), \mathcal{N}(\mu_{src}, \sigma_{src}^2)\} + KL\{\mathcal{N}(prior_{tar}, I), \mathcal{N}(\mu_{tar}, \sigma_{tar}^2)\}$

% 1. model components
% 2. training flow + loss intro
% 3. transfer flow

\section{Experimental setup}
\label{sec:Experiments}

\subsection{Datasets}
\label{ssec:dataset}
% Language mismatch is regarded as one of the most common domain mismatch which has a large impact on SV. 
Two datasets in different languages are used in our experiments of this study. The VoxCeleb 1\&2\cite{nagrani2017voxceleb,chung2018voxceleb2,nagrani2020voxceleb} is one of the most popular large-scale datasets for SV, with the speech data downloaded from YouTube\footnote{https://www.youtube.com}. The data cover various languages, e.g., English and Spanish, but most of them are English. CN-Celeb\cite{fan2020cn} is a Chinese speech dataset with data collected from bilibili\footnote{https://www.bilibili.com}, which is a Chinese video website. The development datasets in VoxCeleb 2 (Vox.2 for short) and CN-Celeb are used for model training, and their information is summarized in Table ~\ref{tab:data}.

The VoxCeleb is regarded as the source domain dataset and the CN-Celeb is the target domain. A speaker embedding extraction model is trained on Vox.2 development set first, which contains 5,994 speakers. The extraction model is evaluated on the test set of VoxCeleb 1 (Vox.(O) for short). The EDITnet's embedding transfer learning is carried out with the data in Table ~\ref{tab:data}. The performance is evaluated on the official test set of CN-Celeb. A total of 3,604,800 utterance pairs are used for the evaluation.
\begin{table}[t]
  \caption{Training data information. The data quantity of CN-Celeb is much smaller than VoxCeleb 2.}
  \label{tab:data}
  \centering
  \begin{tabular}{l|c|c}
    \toprule
                                & \textbf{VoxCeleb 2}       & \textbf{CN-Celeb} \\
    \hline
    Num. of Spk                 & 5,994                      & 800               \\
    Num. of Utt                 & 1,092,009                  & 111,260           \\
    Num. of Hour                & 2360.15                    & 318.12            \\
    Avg. second of Utt          & 7.78                       & 7.72              \\
    Min. second of Utt          & 3.97                       & 0.43              \\
    \bottomrule
  \end{tabular}
\vspace{-2.5mm}
\end{table}
%, without the supervision on any speaker identities.

All data have a sampling rate of $16kHz$ and are transformed into 64-dimension log Mel-filterbanks (FBank) before being fed into the embedding extraction model. The frame window length and hop length used in the experiments are $25ms$ and $10ms$. All acoustic signal processing functions are implemented with the Librosa library\cite{mcfee2015librosa}.

% /exten01/data-set-lijingyu/voxceleb2/train/wav

\subsection{Embedding extraction}
\label{ssec:network}
The embedding extraction model follows the architecture of ECAPA-TDNN\cite{desplanques20_interspeech}, which shows significant performance on VoxCeleb. Specifically, ECAPA-TDNN with $512$ channels (ECAPA-512) is used in this work. For model training, the input data samples are speech segments of 2-second long cropped from random locations of the training utterances. The model parameters are optimized to minimize the Additive Margin Softmax (AM-Soft.)\cite{wang2018additive} cross-entropy on the speaker identities. The dimension of speaker embedding is set as $256$. An equal error rate (EER) of $1.12\%$ is achieved on the Vox.(O), i.e., source domain. The model parameters are fixed in the following EDITnet training.

\subsection{EDITnet training}
\label{ssec:setting}
In each training step, $256$ utterances are randomly sampled from the source domain and target domain respectively. 
% The utterances are cropped into segments with a 2-second duration. From each of the segments, a speaker embedding is obtained by the extraction model. The speaker embeddings are normalized by the mean and standard deviation (std) of the respective domain first and fed into the EDITnet. The mean and std are calculated on each channel of the embeddings from the training data. 
A 2-second long segment is randomly cropped from each utterance. A speaker embedding is obtained by the extraction model from each of the segments. The speaker embeddings are normalized by mean and standard deviation (std) in the first step of the EDITnet. The mean and std are calculated on each channel of the embeddings using the training data from the respective domains. 
The Adam optimizer\cite{DBLP:journals/corr/KingmaB14} is utilized to update the model parameters of EDITnet to minimize the output loss, with a weight decay parameter of $0.001$ and an initial learning rate of $0.001$. The learning rate is declined following a half cosine shape\cite{DBLP:conf/iclr/LoshchilovH17}. The model is trained for $20$ epochs of CN-Celeb, each epoch comprising around $434$ steps. For evaluation, the test utterances and enrollment utterances are divided into segments of 2-second duration, with 1-second overlap between two neighboring segments. The cosine similarity scores between the pairs of test and enrollment segments are calculated and averaged as the score for verification. All experiments are implemented on PyTorch\cite{paszke2019pytorch}.

\section{Results}
\label{sec:Results}

\subsection{Baselines}
\label{ssec:notransfer}
The performance of supervised training on CN-Celeb is evaluated first. The result on ECAPA-512 is given in the first row of Table~\ref{tab:EER_ecapa_cn}. Compared with ECAPA-512's result on Vox.(O) ($EER=1.12\%$), CN-Celeb appears to be a more challenging dataset for SV. Moreover, over-fitting is noted in the training process, as in many other cases of supervised training with an insufficient amount of data. It is considered that, vanilla supervised training is not adequate in this limited data scenario.

With ECAPA-512 trained on Vox.2, directly applied on CN-Celeb, an $EER$ of $17.78\%$ is achieved. This shows the great mismatch between the two domains. The information learned from the source domain is not adequate for speaker discrimination in the target domain, even though VoxCeleb contains a much larger amount of data than CN-Celeb. Supervised domain adaptation is imposed on CN-Celeb by fine-tuning ECAPA-512 with speaker labels in a supervised manner, whose result is shown in Table~\ref{tab:EER_ecapa_cn}. The fine-tuned model performs significantly better than results without domain adaptation.

\begin{table}[t]
  \caption{Performances of ECAPA-512 on the test set of CN-Celeb. The backslash $\backslash$ represents removing a component from the model.}
  \label{tab:EER_ecapa_cn}
  \centering
  \begin{tabular}{clc}
    \toprule
    \textbf{Model training}     &    \textbf{Model transfer}    & \textbf{EER(\%)}  \\
    \midrule
    CN.                                       & None                      & 13.26            \\
    \midrule
    \multirow{2}{*}{\shortstack[l]{Vox.2 }}     & None                                 & 17.78            \\
                                                & CN., supervised                      & \textbf{10.19}   \\
    \midrule
    \multirow{4}{*}{\shortstack[l]{Vox.2 }}     & EDITnet                              & \textbf{12.06}   \\
                                                & EDITnet $\backslash$ pre-norm.        & 12.52            \\
                                                % & EDITnet single output BN             & 12.32            \\
                                                & EDITnet $\backslash$ prior transfer  & 13.65            \\
                                                & EDITnet $\backslash$ cosine loss       & 15.96            \\
    \bottomrule
  \end{tabular}
  \vspace{-2.5mm}
\end{table}

\subsection{Performance of EDITnet}
\label{ssec:editeer}
The proposed EDITnet for domain adaptation gives an $EER$ of $12.06\%$ on the test set of CN-Celeb, which gives a relative improvement of $30\%$ on the non-adapted system. To evaluate the effect of different components in EDITnet, several ablation studies are carried out as described below.

The normalization process can transform the embeddings from two domains into normal\ distribution before feeding them into the CVAE module. The statistics of the embeddings, e.g. mean and std, contain information about the domain. Removing the mean and std differences between input data by normalization is believed to reduce the domain mismatch. Utilizing the normalized embeddings as input, the EDITnet is expected to focus on the information beyond mean and std. When this normalization is absent, the system performance declines slightly to $EER=12.52\%$. It is also noted that this normalization helps stabilize the training of EDITnet.

% At the end of the decoder, the outputs from target and source domain go through two BN layers respectively, i.e. $BN_{tar}$ and $BN_{src}$. They are replaced by one BN layer and the model is retrained. The result is slight worse than the full EDITnet.

An important part of EDITnet is the mean transfer of the latent variable $\mathbf{z}$, i.e., $step\ 7$ in Algorithm~\ref{alg:transfer}. 
% By comparing the two domains' $prior$s as plotted in Fig.~\ref{fig:prior}, the difference between them can be observed. 
The model is retrained and evaluated without the conditional prior in the latent space. In this scenario, all domains are made to approximate a global prior distribution $\mathcal{N}(0, I)$ and essentially no prior transfer takes place in the latent space. The embedding transfer depends only on the encoder and decoder with condition labels. As a result, the performance declines noticeably to $EER=13.65\%$. 

Removing the self-supervised cosine loss in EDITnet training leads to drastic performance degradation, i.e., $EER=15.96\%$. It indicates that the cosine repulsion from self-supervised learning helps to generate discriminative transferred embeddings. In an additional experiment, if the CVAE structure is replaced by two FC layers and the self-supervised cosine loss is retained, the system performance becomes worse. 

% \begin{figure}[t]
%   \centering
%   \includegraphics[width=\linewidth]{figures/prior2.pdf}
%   \caption{The conditional priors for target and source domains.}
%   \label{fig:prior}
% \end{figure}

\subsection{Adaptation on different models}
\label{ssec:othermodels}
To evaluate the generalization ability of EDITnet, two additional embedding extraction models, namely TDNN and SE-ResNet, are experimented. TDNN\cite{snyder2018x} is a classic DNN model for SV, which utilizes 1D-convolution with dilation to capture speech information. SE-ResNet\cite{hu2018squeeze} was first proposed in computer vision. It is utilized in speech analysis by processing the input spectrum (FBANK, MFCC) as a 2D image, and shows good performance in SV\cite{qi2020deep,zhang2021beijing}. The embedding extraction models are trained on Vox.2 with AM-Soft. and the input for the classification layer is utilized as the speaker embeddings. The embedding dimension is set to $256$ for all models.

The experimental results are shown in Table~\ref{tab:EER_all_cn}. It is interesting to note that TDNN and SERes34 have better performance on CN-Celeb than ECAPA-512 without any adaptation, although their EERs on Vox.(O) are worse than ECAPA-512. It suggests that the ECAPA-512 may over-fit on the source domain, thus it can not generalize well on the target domain. EDITnet is compared with several statistics-based transfer methods. The mean and std ($\mu$ and $\sigma$ in Table~\ref{tab:EER_all_cn}) are calculated on the embeddings from the training set. The embeddings from the source domain are normalized to zero mean and unit std by $\mu_{src}, \sigma_{src}$, and transferred to the target domain's distribution by $\mu_{tar}, \sigma_{tar}$.
% Normalizing the embeddings to zero-mean or normal distribution by the target's statistics is called the whitening process, and transferring them into the source's distribution is called the re-color\cite{alam2018speaker}. 
The results show that normalizing the source domain embeddings alleviates the mismatch well, and the transfer process using $\mu_{tar}, \sigma_{tar}$ does not affect much. CORAL\cite{alam2018speaker,sun2016return} utilizes covariance to transfer the target's distribution into source's and shows better performance than previous statistics-based transfer. EDITnet gives the best performance compared with other methods in all three models. It indicates that the proposed transfer model has great ability to capture the feature information, beyond the first or second-order statistics, for alleviating the domain mismatch in the speaker embeddings. There is one point worth mentioning, the EDITnet's performance in SERes34 surpasses the SSDA method ($EER: 9.60\% ~vs. ~ 10.20\%$) proposed by \cite{chen2021self}, which utilizes the ResNet as embedding extractor and have a similar corpus setting, while it requires joint-training with the speaker classification task of the source domain and fine-tuning the whole backbone network.

\begin{table}[t] 
  \caption{The EER($\%$) on the test set of CN-Celeb. The models' performance on the Voxceleb original test set is also provided. $\mathbf{X}$ represents the embeddings extracted from the models.}
  \label{tab:EER_all_cn}
  \centering
  \begin{tabular}{l|ccc}
    \toprule
         &    \multicolumn{3}{c}{\textbf{Model}}  \\
    \hline
    \textbf{transfer method}     &  ECAPA-512    & TDNN         & SE-Res.34        \\
    \hline
    EER on Vox.(O)               &  1.12            & 2.02           & 1.30             \\
    \hline
    None                         &  17.78           & 14.25          & 11.50            \\
    $\mathbf{X}-\mu_{tar}$                &  13.61           & 13.50          & 10.97            \\
    $\mathbf{X}-\mu_{tar}+\mu_{src}$      &  13.63           & 13.58          & 10.82            \\
    $(\mathbf{X}-\mu_{tar})/\sigma_{tar}$    &  13.42        & 13.53          & 10.92            \\
    $(\mathbf{X}-\mu_{tar})/\sigma_{tar}$&  
    \multirow{2}{*}{\shortstack[l]{13.52}}   &\multirow{2}{*}{\shortstack[l]{13.59}}    & \multirow{2}{*}{\shortstack[l]{10.92}}          \\
    $\times\sigma_{src} + \mu_{src}$ & & & \\
    CORAL                        &  12.63           & 13.38          & 10.45            \\      
    \hline
    EDITnet                         &  \textbf{12.06}  & \textbf{12.89} & \textbf{9.60}    \\  
    \bottomrule
  \end{tabular}
  \vspace{-2.5mm}
\end{table}

\section{Conclusions}

This paper describes a lightweight network called EDITnet for fast domain adaption on the embedding level for speaker verification without requiring speaker identity information. The CVAE structure and self-supervised learning strategy are integrated in EDITnet to transfer the speaker embeddings from the target domain into the source domain. Experimental results show that the proposed network is able to mitigate the domain mismatch between datasets of different languages and the transferred embedding demonstrates notable performance improvement against the original embeddings. In addition, EDITnet shows consistent performance gain with different embedding extraction models and outperforms statistics-based transfer methods.

% The domain information is modeled in the latent space of 
\section{Acknowledgements}

Jingyu LI is supported by the Hong Kong PhD Fellowship Scheme of the Hong Kong Research Grants Council.

\bibliographystyle{IEEEtran}

\bibliography{mybib}

% \begin{thebibliography}{9}
% \bibitem[1]{Davis80-COP}
%   S.\ B.\ Davis and P.\ Mermelstein,
%   ``Comparison of parametric representation for monosyllabic word recognition in continuously spoken sentences,''
%   \textit{IEEE Transactions on Acoustics, Speech and Signal Processing}, vol.~28, no.~4, pp.~357--366, 1980.
% \bibitem[2]{Rabiner89-ATO}
%   L.\ R.\ Rabiner,
%   ``A tutorial on hidden Markov models and selected applications in speech recognition,''
%   \textit{Proceedings of the IEEE}, vol.~77, no.~2, pp.~257-286, 1989.
% \bibitem[3]{Hastie09-TEO}
%   T.\ Hastie, R.\ Tibshirani, and J.\ Friedman,
%   \textit{The Elements of Statistical Learning -- Data Mining, Inference, and Prediction}.
%   New York: Springer, 2009.
% \bibitem[4]{YourName17-XXX}
%   F.\ Lastname1, F.\ Lastname2, and F.\ Lastname3,
%   ``Title of your INTERSPEECH 2022 publication,''
%   in \textit{Interspeech 2022 -- 23\textsuperscript{rd} Annual Conference of the International Speech Communication Association, September 18-22, Incheon, Korea, Proceedings, Proceedings}, 2022, pp.~100--104.
% \end{thebibliography}

\end{document}